\documentclass[conference]{IEEEtran}
\IEEEoverridecommandlockouts \usepackage{mathtools}
\usepackage{booktabs}
\usepackage{array}

\usepackage{blkarray}
\usepackage{cite}
\ifCLASSINFOpdf
\fi
\usepackage{amsmath,amsfonts,amsthm} 
\usepackage{amssymb}
\usepackage{bbm}
\usepackage{float}
\usepackage{graphicx}
\usepackage{algorithm}
\usepackage{algpseudocode}
\usepackage{subcaption}
\usepackage{siunitx}
\usepackage{tikz}
\usetikzlibrary{arrows,automata}
\usepackage{multicol}
\usepackage{comment}

\usepackage{pgfplots}
\usepackage{epstopdf}
\usepackage{multirow}
\usepackage[font=scriptsize,skip=5pt]{caption}
\usepackage{soul}
\usepackage{todonotes}
\usepackage{color, colortbl}
\usepackage{array}
\usetikzlibrary{patterns}
\usepackage{balance}

\usepackage{enumitem}
\usepackage{footmisc}
 
\usetikzlibrary{shapes.geometric}

\graphicspath{{images/}}

\newtheorem{remark}{Remark}

\makeatletter
	{\end{proof}%
	\renewcommand{\qedsymbol}{\qedsymbol}}
\makeatother

\DeclareMathAlphabet\mathbfcal{OMS}{cmsy}{b}{n}

\definecolor{Gray}{gray}{0.9}
\definecolor{LightCyan}{rgb}{0.88,1,1}

\algnewcommand{\IIf}[1]{\State\algorithmicif\ #1\ \algorithmicthen}
\algnewcommand{\IElse}[2]{\State\algorithmicelse\ #2\ }
\algnewcommand{\EndIIf}{\unskip\ \algorithmicend\ \algorithmicif}

\IEEEoverridecommandlockouts
\begin{document}
	\setlength{\parskip}{0em}
	\title{
Semantic Text Transmission via Prediction with Small Language Models: Cost-Similarity Trade-off}
	\author{\IEEEauthorblockN{Bhavani A Madhabhavi\IEEEauthorrefmark{1}, Gangadhar Karevvanavar\IEEEauthorrefmark{2}, Rajshekhar V Bhat\IEEEauthorrefmark{3} and Nikolaos Pappas\IEEEauthorrefmark{4}}\\    \IEEEauthorblockA{\IEEEauthorrefmark{1}\IEEEauthorrefmark{2}\IEEEauthorrefmark{3}Indian Institute of Technology Dharwad, Dharwad, India\\
	\IEEEauthorrefmark{4}Department of Computer and Information Science, Link{\"o}ping University, Link{\"o}ping, Sweden\\
		\IEEEauthorrefmark{1}190020011.alum23@iitdh.ac.in, 
		\IEEEauthorrefmark{2}212021007@iitdh.ac.in,
            \IEEEauthorrefmark{3}rajshekhar.bhat@iitdh.ac.in, 
		\IEEEauthorrefmark{4}nikolaos.pappas@liu.se. 
}
}

\maketitle
    \IEEEpeerreviewmaketitle
\pagenumbering{gobble} 
\begin{abstract}
We consider the communication of natural language text from a source to a destination over noiseless and character-erasure channels. We exploit language's inherent correlations and predictability to constrain transmission costs by allowing the destination to predict or complete words with potential dissimilarity with the source text.
Concretely, our objective is to obtain achievable   $(\bar{c}, \bar{s})$ pairs, where  $\bar{c}$ is the average transmission cost at the source and $\bar{s}$ is the average semantic similarity measured via \emph{cosine similarity} between vector embedding of words at the source and those predicted/completed at the destination. We obtain $(\bar{c}, \bar{s})$ pairs for neural language and first-order Markov chain-based small language models (SLM) for prediction, using both a threshold policy that transmits a word if its cosine similarity with that predicted/completed at the destination is below a threshold, and a periodic policy, which transmits words after a specific interval and predicts/completes the words in between, at the destination. We adopt an SLM for word completion. We demonstrate that, when communication occurs over a noiseless channel, the threshold policy achieves a higher $\bar{s}$ for a given $\bar{c}$ than the periodic policy and that the $\bar{s}$ achieved with the neural SLM is greater than or equal to that of the Markov chain-based algorithm for the same $\bar{c}$. The improved performance comes with a higher complexity in terms of time and computing requirements. However, when communication occurs over a character-erasure channel, all prediction algorithms and scheduling policies perform poorly. 
Furthermore, if character-level Huffman coding is used, the required $\bar{c}$ to achieve a given $\bar{s}$ is reduced, but the above observations still apply.

\end{abstract}
 
\section{Introduction}
In the seminal paper, \cite{Shannon1948} Claude Shannon states, ``the fundamental problem of communication is that of reproducing at one point either exactly or approximately a message selected at another point. Frequently the messages have meaning; that is they refer to or are correlated according to some system with certain physical or conceptual entities. These semantic aspects of communication are irrelevant to the engineering problem''. Accordingly, classical communication systems have been developed and optimized, mainly ignoring the semantic aspects of the messages being transmitted, including the correlation between the sequence of messages. This work explores the transmission of natural language text, where consecutive letters and words exhibit correlation. This suggests reducing the transmission cost by allowing the receiver to predict or complete some words. We explore the trade-off between transmission cost and semantic similarity, measured by the average cosine similarity, when predicting or completing words at the receiver. 

Previous works have explored the communication of correlated processes, such as auto-regressive processes, by predicting the realization of the stochastic process at the next instant. Examples include studies on first-order auto-regressive processes \cite{AdityaMahajan, Bhavya}, binary Markov sources \cite{Nikos-GC-19, AoII}, and Wiener processes \cite{WeinerPolyanskiy}. Leveraging the correlated nature of the sources, these works exploit the possibility of predicting the stochastic process in cases where transmission is not carried out. However, in the context of natural language text communication, as also discussed in \cite{Goldsmith, LuoXuewen, XieHuiqiang, QinZhijin}, there has been no exploration of the potential for predicting or completing words to reduce transmission costs.
Nevertheless, prediction in natural language text has been extensively studied outside the context of natural language communication over a channel. For instance, in \cite{BarmanPartha}, next-word prediction in the Assamese language is performed using Long Short-Term Memory (LSTM) and Recurrent Neural Network (RNN) architectures. 
Similar investigations have been conducted in other languages \cite{RadhikaSharma, Shakhovska}. Unlike the above works, we exploit the possibility of predicting natural language text in the context of its transmission over noiseless and erasure channels and reduce the transmission cost. Our work represents one of the first attempts to utilize prediction and word completion to reduce communication costs in communicating natural language text while preserving similarity between the words at the source and those predicted or completed at the destination.

In this work, we investigate the trade-off between average transmission cost and average similarity when employing prediction and word completion at the receiver. This trade-off can be summarized as follows: transmitting every word produces high similarity but also incurs high transmission costs. Not transmitting any word and allowing the receiver to predict eliminates the transmission costs but leads to very low similarity. Our paper makes the following key contributions to address the trade-off mentioned above:
\begin{itemize}[leftmargin=*, noitemsep, topsep=0pt]
    \item 
Considering the plays from the Shakespeare corpus \cite{mit-shakespeare} as the dataset, we determine achievable $(\bar{c}, \bar{s})$ pairs, where $\bar{c}$ represents the average transmission cost at the source, and $\bar{s}$ denotes the average cosine similarity between vector embedding of words at the source and those predicted or completed at the destination. 
The $(\bar{c}, \bar{s})$ pairs are obtained, for 
LSTM-based small language model (LSTM-SLM) and first-order Markov chain-based model (MCM) for word prediction under the following transmission policies: a threshold policy (TP), which does not transmit a word if its cosine similarity with the predicted/completed word at the destination is above a specified threshold, and a periodic policy (PP), which transmits words after a specific word count and predicts/completes the words in between at the destination. In the TP, when the prediction falls below a certain threshold, only the initial characters of a word necessary for the word completion algorithm to complete the word are transmitted. We adopt an RNN-based word completion model. 

    \item  We consider communication over noiseless and character-erasure channels. Our experimental findings reveal that when communication occurs over a character-erasure channel, the achievable $\bar{s}$ for a given $\bar{c}$ is very low under all prediction algorithms and scheduling policies. 
However,   in the noiseless channel case, under both LSTM-SLM and MCM for prediction, TP attains a higher $\bar{s}$ for a given $\bar{c}$ compared to the PP. Moreover, under TP, the $\bar{s}$ achieved with the LSTM-SLM surpasses or equals that of the MCM at the same $\bar{c}$. 
   The improved performance of the LSTM-SLM, which employs a certain number of previous words for prediction, is accompanied by an increase in prediction model complexity, dependent on the number of previous words used. This also impacts performance. 
   Specifically, as the number of previous words used for prediction increases, prediction accuracy improves, reducing the average transmission cost for achieving a given average similarity due to a reduced need for transmitting words. Conversely, increasing the number of words for prediction, increases the average cosine similarity for a given average cost. 
   However, as the number of previous words used for prediction increases, the prediction model becomes more complex, requiring more nodes and longer inference time. In this paper, we provide a numerical characterization of this trade-off.

    \item We also explore the communication of the text using character-level lossless Huffman compression, where each letter in the dataset is assigned a source code based on the Huffman algorithm, determined by the frequency of letter occurrences in the dataset. This approach significantly reduces the average transmission cost $\bar{c}$ for a specified average similarity $\bar{s}$ while maintaining the above performance trends.
\end{itemize}

\section{System Model and Objective}
In this section, we present the source model, possible actions at the source, the channel model, and the considered objective. 
\subsection{System Model}
\subsubsection{Information Source} The source obtains sentences for transmission by getting a word $w_n$ at each discrete-time instant $n\in \{1,2,\ldots\}$, where the word $w_n$ consists of $c_n$ number of characters. 
 
\subsubsection{Possible Actions at the Source} Upon the arrival of the $i^{\text{th}}$ character of the word $w_n$ at the source, the possible actions are: not transmitting the character (indicated by $I_{i,n}=0$) or transmitting it (indicated by $I_{i,n}=1$) for $i \in \{1, 2, \ldots, c_n\}$.   In this work, we assume that a character requires $b$ number of bits for transmission without compression, but with compression, it requires fewer than $b$ bits. 
 
\subsubsection{Communication Channel}
We consider two cases: transmission over (i) a noiseless channel and (ii) a character-erasure channel. In the latter, a character (whether or not compressed) gets replaced by an erasure symbol, $\mathcal{E}$, with probability, $\epsilon$. 

\subsection{Objective}
Let $r(x, y)$ quantify the \emph{semantic similarity} (which will be concretely defined at a later part of the paper) between two words, $x$ and $y$, on a scale of $0$ to $1$, where $1$ indicates exact identity and $0$ represents no similarity. 
 
In this work, our objective is to obtain achievable cost-similarity pairs under 
policy, $\pi\in \{\mathrm{TP}, \mathrm{PP}\}$, for the 
prediction model, $\mathrm{P}\in \{\text{LSTM-SLM}, \text{ MCM}\}$, and character-level compression algorithm, $\mathrm{C}\in \{\text{no compression}, \text{ Huffman coding}\}$.
For a given text corpus consisting of $N$ words, we define the time-averaged cost and similarity, respectively, as
\begin{align}
\bar{c}_{\pi, \mathrm{P}, \mathrm{C}} =    \frac{\sum_{n=1}^N\sum_{i=1}^{c_n} b I_{i,n}}{\sum_{n=1}^N bc_n }, \text{ and } \bar{s}_{\pi, \mathrm{P}, \mathrm{C}} = \frac{1}{N}\sum_{n=1}^N r(w_n, \hat{w}_n),  
\end{align} 
where $\hat{w}_n$ is the estimate of $w_n$ at the destination, after a prediction or word completion, at time slot $n$. 
Concretely, our objective is to obtain
\begin{align}
  &\{\bar{c}_{\mathrm{TP}, \mathrm{P}, \mathrm{C}}, \bar{s}_{\mathrm{TP}, \mathrm{P}, \mathrm{C}},\text{ for all thresholds}\}, \text{ and }\\
  &\{\bar{c}_{\mathrm{PP}, \mathrm{P}, \mathrm{C}}, \bar{s}_{\mathrm{PP}, \mathrm{P}, \mathrm{C}},\text{ for all periods}\}, 
\end{align}
for $\mathrm{P}\in \{\text{LSTM-SLM}, \text{ MCM}\}$ and $\mathrm{C}\in \{\text{no compression}, \text{ Huffman coding}\}$, when communication occurs over noiseless and character-erasure channels.

\section{Solution}
In this section, we derive achievable cost-similarity pairs. We first discuss next-word prediction models, word completion model,  transmission policies, and compression algorithm adopted for communication over noiseless and character-erasure channels. We finally describe the Shakespeare corpus we utilized and present our results. 

\subsection{Prediction Models} 
We employ LSTM-SLM and MCM as next-word prediction models.  Below, we describe these models in detail.

\subsubsection{LSTM-SLM for Next-Word Prediction}
\paragraph{Preliminaries}Let $K$ be the total number of sentences in our dataset, $M_k$ be the number of words in the $k^{\text{th}}$ sentence, where $k\in\{1,2,\ldots, K\}$,  $M_{\text{max}}$ be the length of longest sentence in our training dataset. Let $[w_{i_k}, w_{i_k+1}, \ldots, w_{i_k+M_k}]$ be the $k^{\text{th}}$ sentence, containing $M_k$ words, where $i_k$ is the index of the first word of $k^{\rm th}$ sentence. Let $[t_{i_k}, t_{i_k+1}, \ldots, t_{i_k+M_k}]$ be the tokenized version of the sentence, where each element is a natural number representing a unique word of the dataset. The words are assigned token numbers based on their order of occurrence in the text. The indexing for words in the vocabulary ranges from $1$ to $U$, where $U$ is the total number of unique words present in the entire text. In the case of repeated words, the index already assigned to the word does not change. 

\paragraph{Data Generation}From the tokenized sentence $[t_{i_k}, t_{i_k+1}, \ldots, t_{i_k+M_k}]$, we create all possible $M_k - 1$ sub $\rm n$-gram sentences as follows: $[t_{i_k}, t_{i_k+1}], [t_{i_k}, t_{i_k+1}, t_{i_k+2}], \ldots, [t_{i_k}, t_{i_k+1}, \ldots,$$t_{i_k+M_k}]$. To ensure uniform sentence lengths, we apply zero-padding to obtain the following sequences of length $M_{\rm max}$ for each $k\in \{1,2,\ldots, K\}$:   $[0, \dots,0,t_{i_k}, t_{i_k+1}]$, $[0,\ldots,0,t_{i_k}, t_{i_k+1}, t_{i_k+2}]$ and so on until,  $[0,\ldots,0,t_{i_K}, t_{i_K+1}, \ldots, t_{i_K+M_K}]$.  
When stacked as rows, these vectors form a matrix with entries from natural numbers. The number of rows in this matrix is equal to the total number of samples, denoted as $L= \sum_{k=1}^K (M_k - 1)$, and the number of columns is $M_{\rm max}$. Each row represents a training example, where initial $M_{\rm max}-1$ values are features, and the last value is the label. Specifically, the vector containing initial $M_{\rm max} - 1 $ values will be considered an input sequence. The one-hot representation of the last value will be taken as the label for model training.  
Let $\mathbf{o}_{l}\in \{0,1\}^U$ denote the one-hot vector representation of the label for the $l^{\rm th}$ row (i.e., the last element of the  $l^{\rm th}$ row) of the above matrix for all $l \in \{1, 2, \ldots, L\}$.

\begin{figure}[t]
    \centering
    \begin{center}
\begin{tikzpicture}[scale=0.6,xscale = 0.9, transform shape]
\tikzstyle{box 1}=[draw, minimum height=1.5cm, fill=red!10, text width=4cm,align=center]

\node[box 1, rotate=90, font=\large] (layer2) at (2.5,0) {Embedding Layer };
\node[box 1, rotate = 90, font=\large] (layer3) at (5.5,0) {Bidirectional \\ LSTM Layer};
\node[box 1, rotate = 90, font=\large] (layer4) at (7.5,0) {Dense Layer };

\node[box 1, rotate = 90, font=\large] (layer5) at (9.5,0) {Softmax Layer};

\node[box 1, rotate = 90, font=\large] (layer6) at (12.3,0) {argmax Layer};

\draw[->] (1,0) node[left, above, rotate = 90]{$
\begin{array}{c}
1 \\
2 \\
\vdots \\
L
\end{array}
\begin{bmatrix}
0 & \ldots & 0 & 0 & t_{i_k} \\
0 & \ldots & 0 & t_{i_k} & t_{i_k+1} \\
\vdots & \ddots & \vdots & \vdots & \vdots \\
0 & \ldots  & t_{i_K} & \ldots& t_{i_K+M_K-1}
\end{bmatrix}
$
}  -- (layer2);
\draw[->] (layer2) -- (layer3)node[midway, above]{$\begin{bmatrix}
\mathbf{E_1} \\
\mathbf{E_2} \\
\vdots \\
\mathbf{E_L}
\end{bmatrix}$}; 
\draw[->] (layer3) -- (layer4);
\draw[->] (layer4) -- (layer5);
\draw[->] (layer5) -- (layer6) node[midway, above]{$\begin{bmatrix}
\mathbf{\hat{o}_1} \\
\mathbf{\hat{o}_2} \\
\vdots \\
\mathbf{\hat{o}_L}
\end{bmatrix}$}; 
\draw[->] (layer6) -- (13.5,0) node[right, below, rotate = 90]{$\begin{bmatrix}
\hat{t}_{i_k+1} & \hat{t}_{i_k+2} & \ldots & \hat{t}_{i_K+M_{K}} \\
\end{bmatrix}$};

\end{tikzpicture}

    \end{center}
    \caption{The architecture of the LSTM-SLM for word prediction. }
    \label{fig:LSTM_model}
\end{figure}
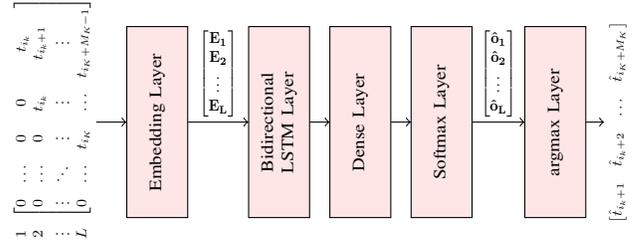

\paragraph{Model Description and Training} The input data is fed into the LSTM-SLM; the model architecture is shown in Fig.~\ref{fig:LSTM_model}. Input sequences pass through an $\rm embedding~layer$, where each token is mapped to a vector representation called an embedding vector of size, $N_{\rm embed}$. Let $\mathbf{E}_l= [\mathbf{e}_{1,l}, \mathbf{e}_{2,l}, \ldots, \mathbf{e}_{M_{\rm max}-1,l}]$ be the embedding layer output, where $\mathbf{e}_{j,l} \in \mathbb{R}^{N_{\rm embed}}$ is the constructed embedding vector for the $j^{\rm th}$ tokenized word of $l^{\rm th}$ input sequence in the training data, for all $j \in \{1, 2, \ldots, M_{\rm max}-1 \}$. Each embedding vector is updated during the training via backpropagation. For each $l^{\rm th}$ input sequence, output of the embedding layer is a matrix of size $(M_{\rm max} - 1) \times N_{\rm embed}$ which further passes through a bidirectional LSTM layer followed by a Dense layer with softmax activation, resulting in an output vector $\mathbf{\hat{o}}_l \in \mathbb{R}^{U}$. Further, argmax of $\mathbf{\hat{o}}_l$ gives the index of the estimated word.
The LSTM-SLM is trained to minimize average cross-entropy loss between $\mathbf{\hat{o}}_l$ and $\mathbf{{o}}_l$ for all $l\in \{1, 2, \ldots, L\}$. 

\paragraph{Communication and Inference} 
Before communication, the trained model is shared with the destination. Initially, a seed text is transmitted to the receiver, and then the trained LSTM-SLM at the receiver is used to predict the next word. Inference follows the same as the training process; first, the seed text is converted into a sequence of tokens and then pre-padded with zeros to match the length of the longest sentence in the training dataset. Subsequently, this sequence is used to predict the token of the next word by passing through several layers of the model. The cosine similarity between the actual word and the predicted word is computed using embedding vectors, and it is explained next.

\begin{remark}
With a slight abuse of notation, let  $\mathbf{e}_{i} \in \mathbb{R}^{N_{\rm embed}}$ and $\mathbf{e}_{j} \in \mathbb{R}^{N_{\rm embed}} $ represent the output of the embedding layer for corresponding tokens, $t_i$ and $t_j$, corresponding to words, $w_i$ and $w_j$, respectively. Then, the cosine similarity between $w_i$ and $w_j$ is given by 
\begin{align}
    r(w_i,w_j) = \cos{\left(\frac{\mathbf{e}_{i}^{\intercal} \mathbf{e}_{j}}{||\mathbf{e}_{i}||_2||\mathbf{e}_{j}||_2}\right)},
\end{align}
where $||\cdot||_2$ is the $2$-norm of a vector. 
Since the vectors capture semantic information in the words, words conveying similar information are expected to have similar embedding vectors. 
\end{remark}

\subsubsection{MCM-based Next-Word Prediction}
Let $u_1, u_2, \ldots, u_U$ represent the unique words in the dataset. We traverse the entire dataset for each word $u_i$ and calculate the relative frequency of $u_j$ appearing after $u_i$, for all $i,j\in \{1,2,\ldots, U\}$. These frequencies are then entered as entries at position $(i, j)$, forming a matrix that serves as a transition probability matrix.
The constructed matrix is shared with the receiver. 
During prediction, when we aim to predict the next word following $u_i$, we choose the value of $j$ corresponding to the highest entry in the transition probability matrix. Then, $u_j$ is taken as the corresponding predicted word.

\subsection{Word Completion Model}
The word completion model is adopted from \cite{word_completion}, which utilizes an RNN to predict the next character based on a fixed number of preceding characters of the sequence of words. During training, we extract a window of fixed length of $100$ characters from the beginning of the training corpus as input features and consider the sequence of $2^{\rm nd}$ to $101^{\rm st}$ characters as the corresponding label. This window is then shifted forward by one character to find another training sequence, and the subsequent $100$ characters after the first character in that window are taken as the label. This process is followed for the entire dataset. During inference, the model takes the preceding $100$ characters as input sequence and estimates the next character. Additionally, the predicted character is added to the previous input, and the window shifts to obtain the next input sequence. This iterative process continues until all characters are either completed or predicted by the model.

\begin{figure*}[t]
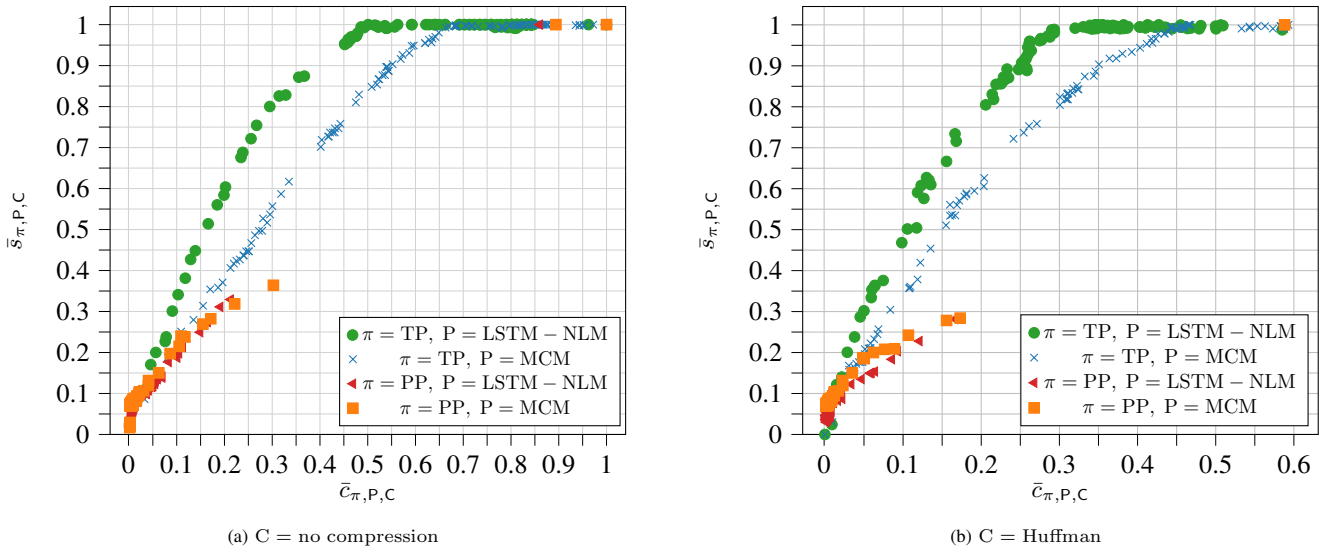

    \begin{subfigure}{.49\textwidth}
    {\input{Plots_12_8_23/wocwoe_right}}
        \caption{$\mathrm{C} = \mathrm{no~compression}$}
    \label{fig:ThresholdVSTRansmitCostNoCompression}
    \end{subfigure}%
    \hspace{2mm}
    \begin{subfigure}{.49\textwidth}
    {\input{Plots_12_8_23/W_enc_wo_era}}
        \caption{$\mathrm{C} = \mathrm{Huffman}$}
        \label{fig:ThresholdVSTRansmitCostCompression}
    \end{subfigure}
    \caption{Achievable average cost-similarity pairs when  communication occurs over a noiseless channel. 
    }
    \label{fig:ThresholdVSTRansmitCostNoErasure}
\end{figure*}

\subsection{Communication Policies}
\subsubsection{Threshold Policy (TP)}
In this policy, the source makes transmission decisions by comparing the receiver's predictions to the actual word at the source, using cosine similarity at a decision instant. 
If the predicted word exists in the vocabulary 
and the cosine similarity exceeds the threshold, no transmission occurs. Otherwise, the source obtains and transmits the characters needed to complete the word's remaining characters, after receiving which, the receiver completes the word. If the completed word is not present in the vocabulary, the entire word is transmitted. If transmission occurs over an erasure channel, some characters may be erased. The word completion model attempts to fill in the erased characters by processing sequence of preceding characters, and adding predicted or non-erased characters into the input sequence to predict next erased character. If the completed word belongs to the vocabulary, the cosine similarity is calculated using the embedding vectors; otherwise, the cosine similarity is considered zero. To run TP, the transmitter needs to know the word predicted at the destination. For a noiseless channel, this can be obtained if the transmitter runs the same model synchronously with the receiver model. For an erasure channel, an acknowledgement from the receiver is needed to inform the transmitter about the received or predicted word.

\subsubsection{Periodic Policy (PP)}
In the PP, words are transmitted at regular intervals/periods of $\tau$ words, i.e., $I_{i,n}=1$ for all $i\in\{1,2,\ldots,c_n\}$  and $n\in \{0,\tau, 2\tau,\ldots\}$ for some period $\tau\in \mathbb{N}$. The words in between the intervals are predicted at the receiver using LSTM-SLM or MCM. Moreover, the PP does not utilize the proximity of its predictions to the actual words, eliminating the need for feedback from the destination.

\begin{remark}
Both in the threshold and periodic policies, when communication occurs over a noiseless channel, all the characters of the transmitted word will be delivered perfectly. The word prediction model can only predict words seen in the training dataset. Since those words have embedding vectors available, the predicted and the actual word at that index can be compared using cosine similarity. Since the transmission is over a noiseless channel, there is no erasure correction required at the receiver. However, if communication occurs over an erasure channel, erased letters are attempted to be recovered using the word completion model. It can happen that a word with some characters erased cannot be completed to achieve a word that belongs to the vocabulary. In such a case, similarity is assigned as zero; if the word does not exist in the vocabulary, there is no concrete method of comparison using the cosine similarity metric. 
\end{remark}

\subsection{Huffman Compression Scheme}
We calculate the relative frequencies of letters, numeric values, and symbols from the text corpus, ${w_1, w_2, \ldots, w_N}$. A codebook is constructed before communication begins, assigning a unique Huffman code to each character based on these relative frequencies. This codebook is transmitted to the receiver prior to the start of communication. According to the policy, when transmission of a word is decided, the sender encodes each letter of the word using the Huffman code, and these Huffman-encoded letters of the word are sent to receiver through either a noiseless or an erasure channel. The receiver decodes letters of the word sent using the codebook.

\begin{figure*}[t]
    \begin{subfigure}{.49\textwidth}
    {\input{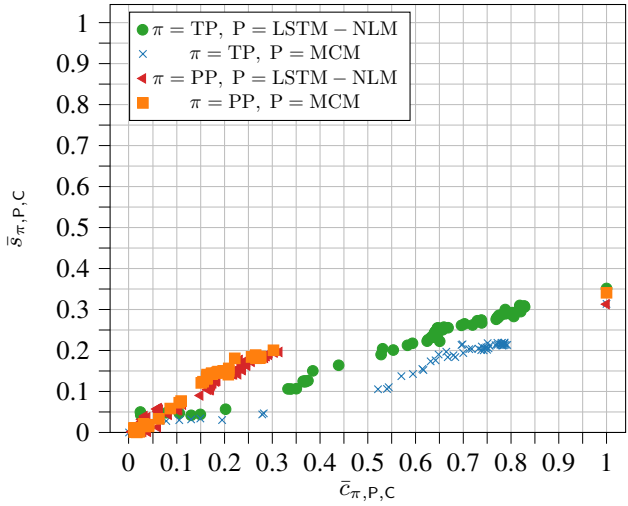}}        \caption{$\mathrm{C} = \mathrm{no~compression}$}
    \label{fig:ThresholdVSTRansmitCostErasureNoCompression}
    \end{subfigure}%
    \hspace{2mm}
    \begin{subfigure}{.49\textwidth}
    {\input{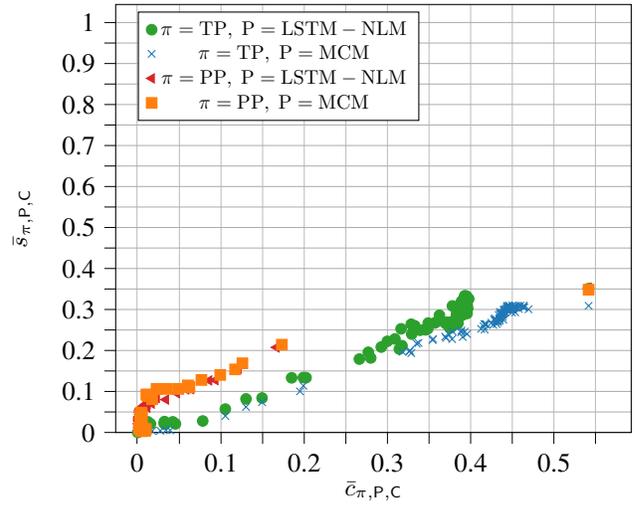}}
        \caption{$\mathrm{C} = \mathrm{Huffman}$}
        \label{fig:ThresholdVSTRansmitCostErasureCompression}
    \end{subfigure}
    \caption{Achievable average cost-similarity pairs when communication occurs over an erasure channel with erasure probability of $0.1$. 
    }
    \label{fig:ThresholdVSTRansmitCostErasure}
\end{figure*}

\section{Numerical Results}

\begin{figure}[t]    
     {
\begin{tikzpicture}

\definecolor{crimson2143940}{RGB}{214,39,40}
\definecolor{darkgray176}{RGB}{176,176,176}
\definecolor{darkorange25512714}{RGB}{255,127,14}
\definecolor{lightgray211}{RGB}{211,211,211}
\definecolor{forestgreen4416044}{RGB}{44,160,44}
\definecolor{mediumpurple148103189}{RGB}{148,103,189}
\definecolor{sienna1408675}{RGB}{140,86,75}
\definecolor{steelblue31119180}{RGB}{31,119,180}

\begin{axis}[
tick align=outside,
tick pos=left,
x grid style={lightgray211},
xlabel={\scriptsize The number of previous words used for prediction, $L$},
xmajorgrids,
xmin=-2.9, xmax=104.9,
xminorgrids,
xtick style={color=black},
y grid style={lightgray211},
ylabel={\scriptsize Average transmission cost, $\bar{c}$},
ymajorgrids,
ymin=0, ymax=1,
yminorgrids,
ytick style={color=black},
xtick={2,10,25,50,75,100},
xticklabel style={rotate=0},
xticklabels={2, 10, 25, 50, 75, 100},
legend style={at={(1, 1)},anchor=north east},
legend style={nodes={scale=0.7, transform shape}}
]
\addplot [semithick, steelblue31119180, mark=*, mark size=3, mark options={solid}]
table {%
2 0.411492122335496
10 0.317886932344764
25 0.209453197405005
50 0.132530120481928
75 0.0843373493975904
100 0
};
\addlegendentry{Threshold = $0.1$}
\addplot [semithick, darkorange25512714, mark=square*, mark size=3, mark options={solid}]
table {%
2 0.707136237256719
10 0.675625579240037
25 0.541241890639481
50 0.438368860055607
75 0.381835032437442
100 0.253012048192771
};
\addlegendentry{Threshold = $0.2$}
\addplot [semithick, forestgreen4416044, mark=x, mark size=3, mark options={solid}]
table {%
2 0.85356811862836
10 0.764596848934198
25 0.705282669138091
50 0.588507877664504
75 0.484708063021316
100 0.428174235403151
};
\addlegendentry{Threshold = $0.3$}
\addplot [semithick, crimson2143940, mark=diamond*, mark size=3, mark options={solid}]
table {%
2 0.927710843373494
10 0.860055607043559
25 0.789620018535681
50 0.651529193697868
75 0.573679332715477
100 0.515291936978684
};
\addlegendentry{Threshold = $0.4$}
\addplot [semithick, mediumpurple148103189, mark=pentagon*, mark size=3, mark options={solid}]
table {%
2 0.972196478220575
10 0.927710843373494
25 0.827618164967563
50 0.727525486561631
75 0.624652455977757
100 0.545875810936052
};
\addlegendentry{Threshold = $0.6$}
\addplot [semithick, sienna1408675, mark=star, mark size=3, mark options={solid}]
table {%
2 1
10 0.962001853568119
25 0.849860982391103
50 0.749768303985171
75 0.64967562557924
100 0.572752548656163
};
\addlegendentry{Threshold = $1.0$}
\end{axis}

\end{tikzpicture}} 
    \caption{Impact of varying the number of previous words used for prediction on the transmission cost, with LSTM-SLM for TP, for transmission over a noiseless channel.}
    \label{fig:VarySeedTextLength}
\end{figure}
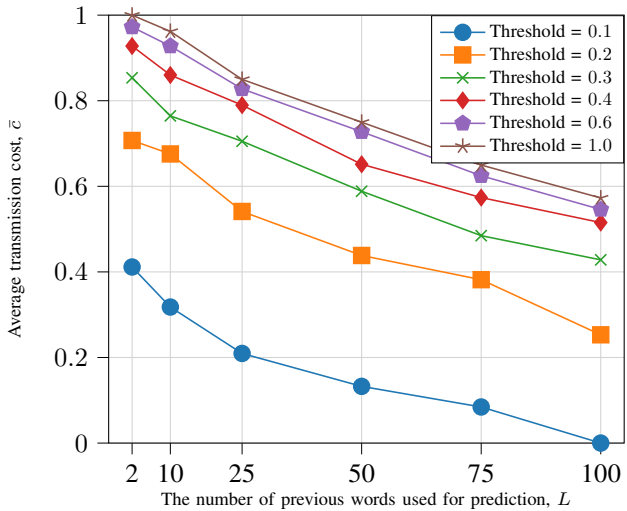

We utilize the plays from the Shakespeare corpus, a commonly employed dataset in  natural language processing tasks. The corpus comprises over a million lines of text, representing the dialogues within the plays and containing a total of $54$k complete sentences. We preprocess the dataset by excluding certain insignificant words to enhance the prediction task. The remaining unique words are compiled into a vocabulary dictionary.  For training, we utilize $75$k lines of text, equivalent to $36$k complete sentences, and test on $100$ complete sentences. 
Below, we present our results. 

In Fig.~\ref{fig:ThresholdVSTRansmitCostNoErasure},  and Fig.~\ref{fig:ThresholdVSTRansmitCostErasure}, we present the performance of different prediction models (LSTM-SLM and MCM) and scheduling policies (TP and PP). We make following observations: Firstly, when communicating over a noiseless channel, TP results in higher average similarity for a given average cost than PP, regardless of compression and prediction models. With TP, LSTM-SLM achieves higher average similarity than MCM, whether compression is utilized or not. This is because the LSTM-SLM-based prediction algorithm outperforms the MCM-based prediction algorithm by making more accurate predictions, as it leverages a larger context of previous words compared to the MCM-based prediction, which relies on only the preceding word.  This improved prediction enables TP to make better-informed decisions compared to PP, as TP utilizes cosine similarity for decision-making. Secondly, when communication occurs over a character-erasure channel, the average achievable similarity for a given average cost under TP and PP, with both LSTM-SLM and MCM, is very low and more or less similar. This is because erasures degrade the performance of both prediction and word completion algorithms. Finally, we observe that compression via Huffman coding reduces the average transmission cost required to achieve a cosine similarity of one.  Moreover, the performance of both policies are worse in the erasure channel case, compared to noiseless channel, and the performance gap between LSTM-SLM and MCM-based prediction algorithm with TP is smaller in the erasure channel case.  
This is because, during communication over an erasure channel, when a word or a letter is transmitted, one or more letters may be erased. In such cases, the  word completion algorithm fills in the erased letters and completes the word. However, the  word completion algorithm may predict incorrect letters in place of erased characters, leading to decrease in  cosine similarity and impacting subsequent predictions.

From Fig.~\ref{fig:ThresholdVSTRansmitCostErasure}, we observe that the PP for low costs has a higher similarity than the TP for both LSTM-SLM and MCM, when communication occurred over erasure channel. In the TP, if the similarity score for predicted word does not exceed the threshold, the initial characters of the word at the source, which are sufficient for the word completion algorithm to reconstruct the word, are transmitted over the erasure channel. The  same word completion model at the receiver tries to correct the erasures by predicting the characters in those positions. However, multiple errors could produce a cascading effect and degrade the overall similarity, as mentioned above. However, in the PP, words need to be transmitted after a certain period of time, during which words are predicted at the receiver,  this leads to a lower average cost.  The words  predicted at the receiver do not face any erasures; they always belong to the vocabulary, allowing cosine similarity to be calculated.  Thus, the performance of the PP is slightly better than that of the TP. Both policies exhibit similar performance when the average transmission cost is close to zero or close to one, when no compression is employed. However, LSTM-SLM performs better in the TP, when the cost deviates from zero to one. The average similarity between the predicted and actual words is close to zero when the average cost is close zero and close to one when the average cost is close to one.

Given the superior performance of the LSTM-SLM-based prediction algorithm with the TP, we study it further in Fig.~\ref{fig:VarySeedTextLength} and Table~\ref{tab:table1}. In Fig.~\ref{fig:VarySeedTextLength}, we examine how the transmission cost varies with the number of previous words used for predicting the next word. As the number of previous words used for prediction increases, the prediction performance of the model improves in terms of cosine similarity, as it utilizes the information of previous words to predict next word. Hence, more and more predicted words exceed the threshold in the TP, leading to decrease in average transmission cost.    
 
Table~\ref{tab:table1} compares the average transmission cost in the TP with word prediction and word completion models, across different thresholds and similarities. Solely relying on the word prediction model necessitates complete word transmission when the threshold is not met, which resulting in increased costs. Combining prediction and word completion algorithms further reduces the transmission cost. With the word completion model, transmitting fewer characters is sufficient, as the receiver can predict the subsequent characters.
We observe that, in word prediction LSTM-SLM yields a lower transmission cost than MCM for a given average cosine similarity. However, LSTM-SLM incurs more delay (about $80$ times more) in predicting the next word compared to MCM, which is a simpler model. The average word prediction time, $T_{\rm avg}$, is calculated  for the inference performed on the Nvidia T4 GPU.

 In Table~\ref{tab:table2}, we provide the results showing the percentage contribution of the word prediction and word completion algorithms to reducing the average transmission cost in achieve certain average similarity using the TP. This analysis assumes no compression and communication over a noiseless channel. We observe that as the threshold increases, the majority of cost reduction comes from the word completion model. This is because the predictions made by the word prediction model fail to surpass the higher threshold values; consequently, the word completion model takes over. In this scenario, the initial words are transmitted, and the word completion algorithm successfully completes the words, thereby increasing the overall similarity, with lower transmission cost. 

 \balance 
\section{Conclusion}
We considered communication of natural language text over noiseless and character-erasure channels by exploiting the correlations and predictability of language to constrain transmission costs by allowing the receiver to predict or complete words. We numerically characterized achievable average cost-similarity pairs under different word prediction and scheduling algorithms.
In conclusion, our work shows that the threshold policy outperforms the periodic policy when communication takes place over a noiseless channel, achieving a higher $\bar{s}$ for a given $\bar{c}$. Additionally, we find that a small language model yields an equal or greater $\bar{s}$ compared to the Markov chain-based model for the same $\bar{c}$. This enhanced performance comes at the cost of increased prediction model complexity. However, when communication occurs over a character-erasure channel, all prediction models and scheduling policies exhibit poor performance.
Further, compression via Huffman coding reduces the transmission cost to achieve a certain similarity.  

\begin{table}[t]
\vspace{0.3cm}
    \caption{The comparison of average transmission cost in the TP for word prediction and word completion algorithms, when no compression is employed and communication occurs over a noiseless channel. $T$ is the threshold, $L$ is the number of previous words, and  $T_{\rm avg}$ is the average time required for a word prediction.}
    \centering
  \begin{tabular}{|p{1.65cm}|p{1.15cm}|p{1.15cm}|p{1.10cm}|p{1.35cm}| }
    \hline
        \text{Prediction, $\mathrm{P}$}   &   \text{Word}   & \text{$T = 0.75$} & \text{$T = 1$} & \text{$T_{\rm avg}$ (in } \\ 
        {} & {\text{Completion}} & {$L = 100$} & {$L = 100$} & {\text{milliseconds})} \\
        {} & {} & {$\bar{s}$ = 0.93} & {$\bar{s}$ = 1} & {} \\[4pt]
    \hline
    \hline 
        LSTM-SLM         &   -   & $0.814$      & $0.838$ & $7.8$\\[4pt]
    \hline
        MCM &   -   & $0.963$  & $0.987$ & $0.1$\\[4pt]
    \hline
        LSTM-SLM        &   RNN   & $0.638$      & $0.643$ & $7.8$\\[4pt]
    \hline
        MCM &   RNN   & $0.749$  & $0.776$ & $0.1$\\[4pt]
    \hline

    \end{tabular}
    \label{tab:table1}
\end{table}

\begin{table}[t]
    \caption{The contribution of word prediction (WP) and word completion (WC) algorithms to reducing the average cost in the TP to achieve certain average similarities.}
    \centering
      \renewcommand{\arraystretch}{0.6}    
  \begin{tabular}
  {|p{1.5cm}|p{1.55cm}|p{1.35cm}|p{1.15cm}|p{1.10cm}| }
    \hline
        \text{\% reduction}   &   \text{\% reduction}   & \text{Average} & \text{Threshold,} & \text{Average} \\ 
        {\text{in cost by WP}} & {\text{in cost by WC}} & {\text{similarity,} $\bar{s}$} & {$T$} & {\text{cost,} $\bar{c}$} \\ [4pt]
    \hline
    \hline 
        $74$   &   $26$   & $0.35$  & $0.1$ & $0.12$
        \\[4pt]
    \hline
        $20$   &   $80$   & $0.9$   & $0.3$ & $0.33$ \\[4pt]
    \hline
        $10$   &   $90$   & $0.95$ & $0.6$ & $0.45$\\[4pt]
    \hline
        $5$    &   $95$   & $0.98$  & $0.9$ & $0.89$\\[4pt]
    \hline
        $2$    &   $98$   & $1$     & $1$   & $1$\\
    \hline
    \end{tabular}
    \label{tab:table2}
\end{table}

\bibliographystyle{ieeetr}
\bibliography{references}

\end{document}